# On the Relationship Between Magnetic Expansion Factor and Observed Speed of the Solar Wind from Coronal Pseudostreamers

Samantha Wallace,[1,2] C. Nick Arge,[2] Nicholeen Viall,[2] and Ylva Pihlström[1,*]

[1] Department of Physics and Astronomy
University of New Mexico
210 Yale Blvd NE
Albuquerque, NM 87106, USA
[2] NASA Goddard Space Flight Center
8800 Greenbelt Rd
Greenbelt, MD 20771, USA

## ABSTRACT

For the past 30+ years, the magnetic expansion factor ($f_s$) has been used in empirical relationships to predict solar wind speed ($v_{obs}$) at 1 AU, based on an inverse relationship between these two quantities. Coronal unipolar streamers (*i.e.* pseudostreamers) undergo limited field line expansion, resulting in $f_s$-dependent relationships to predict fast wind associated with these structures. However, case studies have shown that in situ observed pseudostreamer solar wind was much slower than that derived with $f_s$. To investigate this further, we conduct a statistical analysis to determine if $f_s$ and $v_{obs}$ are inversely correlated for a large sample of periods when pseudostreamer wind was observed at multiple 1 AU spacecraft (*i.e.* ACE, STEREO-A/B). We use the Wang-Sheeley-Arge (WSA) model driven by Air Force Data Assimilative Photospheric Flux Transport (ADAPT) photospheric field maps to identify 38 periods when spacecraft observe pseudostreamer wind. We compare the expansion factor of the last open field lines on either side of a pseudostreamer cusp with the corresponding in situ measured solar wind speed. We find only slow wind ($v_{obs} < 500$ km/s) is associated with pseudostreamers, and that there is not a significant correlation between $f_s$ and $v_{obs}$ for these field lines. This suggests that field lines near the open-closed boundary of pseudostreamers are not subject to the steady-state acceleration along continuously open flux tubes assumed in the $f_s$–$v_{obs}$ relationship. In general, dynamics at the boundary between open and closed field lines such as interchange reconnection will invalidate the steady-state assumptions of this relationship.

*Keywords:* solar wind — Sun: magnetic fields — Sun: corona

## 1. INTRODUCTION

The solar wind is a result of the supersonic expansion of hot (T $\sim 10^6$ K) plasma and magnetic field in the solar corona. This highly ionized plasma with coronal mangetic fields frozen into it flows out into the heliosphere, with observed speeds ($v_{obs}$) ranging between ~250-750 km/s (Feldman et al. 1978). This outflow can be approximated in models as originating from regions on the Sun that have largely unipolar magnetic field and that are magnetically "open" (*i.e.* flux tubes with only one footpoint connected to the Sun), or coronal holes. Coronal holes have lower temperature and density relative to the background corona, and are thus identified in remote coronal observations by their reduced X-ray and extreme ultraviolet (EUV) emission. It is important to note that the boundaries of coronal holes observed remotely may not be precisely where the magnetic open-closed boundaries are located (de Toma et al. 2005).

* Y.M. Pihlström is also an Adjunct Astronomer at the National Radio Astronomy Observatory

How the solar wind is accelerated is an area of active research, in which theory and empirical relationships are heavily relied upon. Thirty years ago, Wang & Sheeley (1990) discovered an inverse relationship between solar wind speed (as measured by spacecraft near Earth) and the coronal field line expansion at the location that the observed solar wind emerged from. Using a magnetostatic potential field source surface (PFSS) model (Schatten et al. 1969; Altschuler & Newkirk 1969; Wang & Sheeley 1992), they extrapolated the coronal field out to 2.5 $R_\odot$ (Hoeksema et al. 1983) from photospheric field observations at 1 $R_\odot$ and quantified the rate of inferred expansion of a coronal magnetic flux tube compared to a $R^{-2}$ drop off with the following equation:

$$f_s = \left(\frac{R_{ph}}{R_{ss}}\right)^2 \left(\frac{B_{ph}}{B_{ss}}\right) \tag{1}$$

where $B_{ph}$ and $B_{ss}$ are the field strengths along each flux tube at the photosphere ($R_{\rm ph} = 1\ R_\odot$) and source surface ($R_{\rm ss} = 2.5\ R_\odot$) respectively (Wang & Sheeley 1992). By tracing model-derived magnetic field lines from the Earth back to the Sun, they found that fast solar wind ($v_{obs} > 500$ km/s) is correlated with the centers of coronal holes where $f_s$ is small, while slow solar wind ($v_{obs} < 500$ km/s) originates from coronal hole (CH) boundaries where $f_s$ is large. This discovery was an important breakthrough, as it provided the heliophysics community with a way to both predict and forecast the solar wind (Arge & Pizzo 2000; Pizzo et al. 2011).

While the practical importance of the inverse relationship between observed solar wind speed and expansion factor is without dispute (Sheeley 2017), the physical interpretation and relevance behind it has been debated ever since. In wave-turbulence driven (WTD) acceleration theories, energy deposited into the corona is a function of a flux tube's radius. Thus, differences in observed solar wind speed are attributed to the rate of flux-tube expansion in the low corona, implying a physical connection between $f_s$ and $v_{obs}$. This is supported by quantitative theoretical arguments and modeling using steady-state, continuously open flux tubes (Wang et al. 1996; Cranmer et al. 2007). In addition to speed, steady-state models are generally able to replicate long-term empirical trends between $f_s$ and solar wind density, Alfvénicity, and charge state (Wang et al. 2003; Cranmer et al. 2007; Wang et al. 2009; Cranmer 2010).

In contrast, reconnection/loop-opening (RLO) theories argue that the magnetic reconnection of open field lines with closed magnetic loops imparts both the energy and mass flux into the overlying corona needed to obtain the terminal speed of the solar wind (Fisk 2003). This theory assumes a fixed energy deposition at the coronal **base,** as opposed to depositing energy per unit volume as a function of radius as in WTD and flux-tube expansion based models. Fisk (2003) argued that slow wind emerges from coronal hole boundaries where there is access to larger, denser closed loops from within streamers and fast wind emerges from deep inside coronal holes where there is only access to very small loops. Thus, he asserted that the existence of the $f_s$–$v_{obs}$ relationship is simply a coincidence due to the magnetic topology of closed magnetic field. Reconnection/loop-opening theories are also generally able to reproduce long term trends between observed speed, density, and charge states (Schwadron et al. 1999; Fisk 2003). These theories additionally provide an explanation for the First Ionization Potential (FIP) enhancement observed in some slow wind observations (Geiss et al. 1995; Zurbuchen et al. 1998). However, there is a growing understanding in the Heliophysics community that WTD and RLO acceleration theories are not mutually exclusive, and so determining which theory plays a dominant role and under what circumstances is essential to progress (Cranmer 2009; Viall & Borovsky 2020).

Similarly, Riley et al. (2001) developed an alternative empirical relationship that predicts solar wind speed based on the minimum angular separation between the solar wind source (*i.e.* open field line footpoint) and the nearest open-closed boundary at 1 $R_\odot$, or "coronal hole boundary distance" (DCHB or $\theta_b$). In this relationship when a field line has a small $\theta_b$, its footpoint is close to the open-closed boundary and the solar wind speed is slow. While the empirical relationship between $\theta_b$ and $v_{obs}$ is not inherently physical, a major difference between this relationship and $f_s$–$v_{obs}$ relationships is that it could be a proxy for magnetic reconnection. It also does not constrain the magnetic field lines to being continuously open.

Periods for which $f_s$-dependent empirical relationships have performed poorly are when the in-situ observed solar wind was formed at a coronal unipolar streamer. Otherwise known as pseudostreamers, these solar magnetic structures differ from their dipolar counterpart (*i.e.* helmet streamers) in that they form from two converging coronal hole boundaries of the *same* polarity and therefore do not form a current sheet. Instead, these field lines converge above the cusp (*i.e.* X-point), and limit the expansion of the underlying closed field (Wang et al. 2012). Thus, pseudostreamers in theory would have smaller expansion factors than helmet streamers, leading Wang et al. (2007) to originally postulate that pseudostreamer wind was fast. However, Riley & Luhmann (2012) used a global MHD coronal model to identify a period when ACE was well-positioned to observe the solar wind that emerged from a pseudostreamer. They found that the observed solar wind was slow, yet the predicted speed based on the original Wang-Sheeley (WS) relationship was fast due to the low expansion factors associated with this structure. This work was expounded on in Riley et al. (2015) to test the use of both $f_s$ and $\theta_b$ to predict solar wind speed. They found that on average, empirical relationships relying either solely or mostly on $\theta_b$ outperform the original WS $f_s$–$v_{obs}$ relationship, especially when pseudostreamers are present. They conclude that $\theta_b$ predicts solar wind speed better than $f_s$, and suggest that their findings may rule out a causal relationship between solar wind speed and $f_s$ (or at least relegate it to a minor role). However, both studies only investigated in detail one particular Carrington rotation where a well-defined pseudostreamer was observed (CR 2060).

In this study, we build upon the work of Riley & Luhmann (2012) and Riley et al. (2015), and investigate the relationship between expansion factor ($f_s$, as originally defined by Wang & Sheeley 1990) and observed solar wind speed for several periods when multiple 1 AU spacecraft (ACE, STEREO-A, STEREO-B) observe pseudostreamer wind. We exploit the rigorous capabilities of the Wang-Sheeley-Arge (WSA) model (Arge & Pizzo 2000; Arge et al. 2003, 2004) coupled with Air Force Data Assimilative Photospheric Flux Transport (ADAPT: Arge et al. 2010, 2011, 2013; Hickmann et al. 2015) photospheric field maps and develop a methodology to determine the precise source regions of the in situ observed solar wind. This methodology is used to identify periods when the observed solar wind emerged from the last two model-derived open field lines converging at a pseudostreamer X-point, where the spacecraft connectivity changes from one coronal hole boundary to another of like-polarity. The individual model-derived solar wind parcel that emerges from each field line is then propagated outward to an observing spacecraft, and the model-predicted arrival time is used to record the in situ observed solar wind speed. This observed speed is then compared to the corresponding expansion factor. We perform a statistical analysis over all identified field lines to examine whether expansion factor and observed solar wind speed are correlated for periods when pseudostreamer wind is observed.

This paper is structured as follows. Section 2 gives an overview of the ADAPT-WSA model, and outlines the methodology used to identify periods when spacecraft observe pseudostreamer wind. The results are presented in Section 3. They are then discussed in the context of other empirical solar wind speed relationships and solar wind formation theories in Section 4, and summarized in the final section.

## 2. IDENTIFYING IN SITU OBSERVED PSEUDOSTREAMERS WITH ADAPT-WSA

In this section, the ADAPT-WSA model is summarized and the methodology we developed to identify periods when spacecraft sample the solar wind that emerged from pseudostreamers is outlined in detail. While this methodology is applied to pseudostreamers in this work, it can also be used to investigate solar wind that originates from other sources (*e.g.* helmet streamers, coronal holes, plasma originating from or near active regions).

### 2.1. *The ADAPT-WSA model*

The WSA model is a combined empirical-and physics-based model that is an improved version of the original WS model (Wang & Sheeley 1992, 1995). WSA relies on input global photospheric field maps assembled from full-disk observations of the solar photospheric magnetic field (*i.e.* magnetograms). These maps are constructed in a variety of ways representing either a time history of central meridian evolution over a Carrington rotation (*i.e.* diachronic) or, more preferably, one moment in time (*i.e.* synchronic). Given the current lack of far-side solar magnetic field observations and poor observations of the poles, global synchronic representations of the photospheric magnetic field are only possible through flux-transport models (Worden & Harvey 2000; Schrijver & De Rosa 2003). In this work, we use global synchronic photospheric field maps generated by the ADAPT model. The ADAPT model utilizes magnetic flux transport based on the Worden and Harvey (2000) model to account for differential rotation, along with meridional and supergranulation flows, when observational data are not available. In addition, ADAPT incorporates new magnetogram input using the ensemble least-squares data assimilation technique accounting for both model and data uncertainties as the maps are generated (Hickmann et al. 2015). For example, ADAPT heavily weights observations taken near disk center where magnetograms are most reliable, while the model specification of the field is generally given more weight near the limbs where observations are the least reliable. ADAPT produces an ensemble of maps (or realizations) for any given moment in time that ideally represents the uncertainty in the global photospheric magnetic field distribution.

Using ADAPT maps (*e.g.* Figure 1b) as input, WSA derives the coronal field using a coupled set of potential-field type models. The first is a traditional PFSS model, which determines the coronal field out to the source surface height. The traditional source surface height of 2.5 $R_\odot$ (Hoeksema et al. 1983) is used in this study because it has been shown in prior work to produce good agreement between WSA-derived open flux and that derived from Helium and EUV coronal hole observations over nearly two solar cycles (Wallace et al. 2019) with the same set of Vector Spectromagnetograph (VSM: Henney et al. 2009) magnetograms used in this study. The output of the PFSS model serves as input to the Schatten Current Sheet (SCS) model (Schatten 1971), which provides a more realistic magnetic field topology of the upper corona. Although this solution extends out to infinity, WSA uses a portion the coronal field solution that terminates at an outer boundary radius set by the user (5 $R_\odot$ for this work). The following empirical velocity relationship is then used to determine the solar wind speed of each magnetic field line at the outer boundary:

$$V(f_s, \theta_b) = 285 + \frac{685}{(1+f_s)^{2/9}}\{1 - 0.8e^{-(\theta_b/2)^2}\}^3 \tag{2}$$

which is a function of both expansion factor and the minimum angular separation between an open field line footpoint at 1 $R_\odot$ and the nearest open-closed boundary. Instead of "back-mapping" a spacecraft to the outer boundary of the model, WSA propagates solar wind parcels outward from the endpoints of each field line located at 5 $R_\odot$ to an observing spacecraft (*i.e.* ACE, STEREO-A &B in this study). Stream interactions are accounted for in the solar wind propagation using a simple 1-D modified kinematic model, which prevents fast streams from bypassing slow ones (Arge et al. 2004). When coupled with ADAPT, WSA derives an ensemble of 12 solutions, each representing the global state of the coronal field and connectivity from a spacecraft to 1 $R_\odot$ for a given moment in time. The best realization is then determined by comparing the model-derived and observed interplanetary magnetic field (IMF) and solar wind speed.

Since WSA uses a magnetostatic coronal model, it is not possible for the model to capture the Sun's time dependent phenomena associated with the opening and closing of magnetic flux. While we can account for time-dependent photospheric phenomena with ADAPT, WSA only derives the magnetic connectivity between an observing spacecraft and model-derived field lines that are open. Similarly, WSA cannot provide information regarding how long a particular field line has been open. Therefore, when the model predicts that a spacecraft measured plasma near a closed-flux system (*i.e.* the open-closed boundary), the two physical scenarios that are possible are 1) the plasma originated from that particular open field line, or 2) the plasma originated on closed field that was recently opened via interchange reconnection (IR), whereas ADAPT-WSA cannot make the distinction between the two possible scenarios.

### 2.2. *Methodology*

Figures 1−3 present one period when solar wind that emerged from a pseudostreamer (highlighted in green in all three figures) is observed by STEREO-A, and are used to illustrate the methodology of this work. We first use the ADAPT-WSA model to derive the global coronal field (Figure 1c), as well as the connectivity between the projection of STEREO-A's location at 5 $R_\odot$ and the open field footpoints at 1 $R_\odot$ (Figure 1a). Throughout Figure 1, dates labeled in red above the white/red cross hairs (*i.e.* sub-satellite points, see Figure 1 for definition) correspond to when and where the solar wind *left* the Sun at 5 $R_\odot$, as opposed to when it *arrived* at STEREO-A. Similarly, black lines extend from the sub-satellite track to 1 $R_\odot$, revealing the model-derived source regions of the solar wind that ultimately was observed at STEREO-A. The connectivity plots (*e.g.* Figure 1a) are used to identify periods when the in situ observed solar wind was formed at a pseudostreamer, by searching for instances when the spacecraft connectivity (*i.e.* black lines) changes from one coronal hole boundary to another of the same polarity (indicated in grayscale). Figure 1a reveals that on July 12th, 2014, STEREO-A was magnetically connected to the boundaries of two mid-latitude coronal holes of positive polarity, labeled with green lines that connect the sub-satellite track (*i.e.* white cross hairs) to either coronal hole. Thus, on July 12th the solar wind emerged from this specific location as derived by ADAPT-WSA, and propagated outward to eventually be observed at STEREO-A ∼4−6 days later. Figure 1c further confirms that the STEREO-A sub-satellite track is embedded entirely in positive field at 5 $R_\odot$ for the days surrounding July 12th.

Before investigating the specific field lines of each pseudostreamer that are connected to a spacecraft, the best ADAPT-WSA solution of the 12 must be selected for each period of interest. To do so, we compare the WSA-derived solar wind speed and IMF polarity for all 12 realizations with observations from ∼4−6 days after the solar wind left the Sun (*i.e.* approximate travel time of solar wind to 1 AU). The best realization for the pseudostreamer identified in Figure 1 is shown in Figure 2, which compares the model-derived (blue) and STEREO-A observed (black) solar wind speed and IMF for a portion of CR 2152 (approximately two weeks). For the days surrounding the estimated time-of-arrival of the pseudostreamer wind at STEREO-A (*i.e.* July 17−19th), the model-derived solar wind speed and IMF agree well with that observed at STEREO-A. While there are other instances during this CR where the model did not accurately derive the solar wind speed (*i.e.* 21 Jul, 25 − 30 Jul), the model-derived parameters for the days surrounding the period of interest agree well. This gives us confidence in the model-derived connectivity between solar wind parcels propagated out from pseudostreamer field lines and STEREO-A. Since ADAPT produces photospheric field solutions for a fixed moment in time, maps used within a few days of each period of interest generally produce the most realistic solutions, whereas a different methodology would be necessary to obtain good agreement with observations over an entire CR. Plots such as Figure 2 are generated for all 12 realizations and used initially to quickly determine the most realistic model solution for the time period of interest. This plot is revisited in greater detail later in this section.

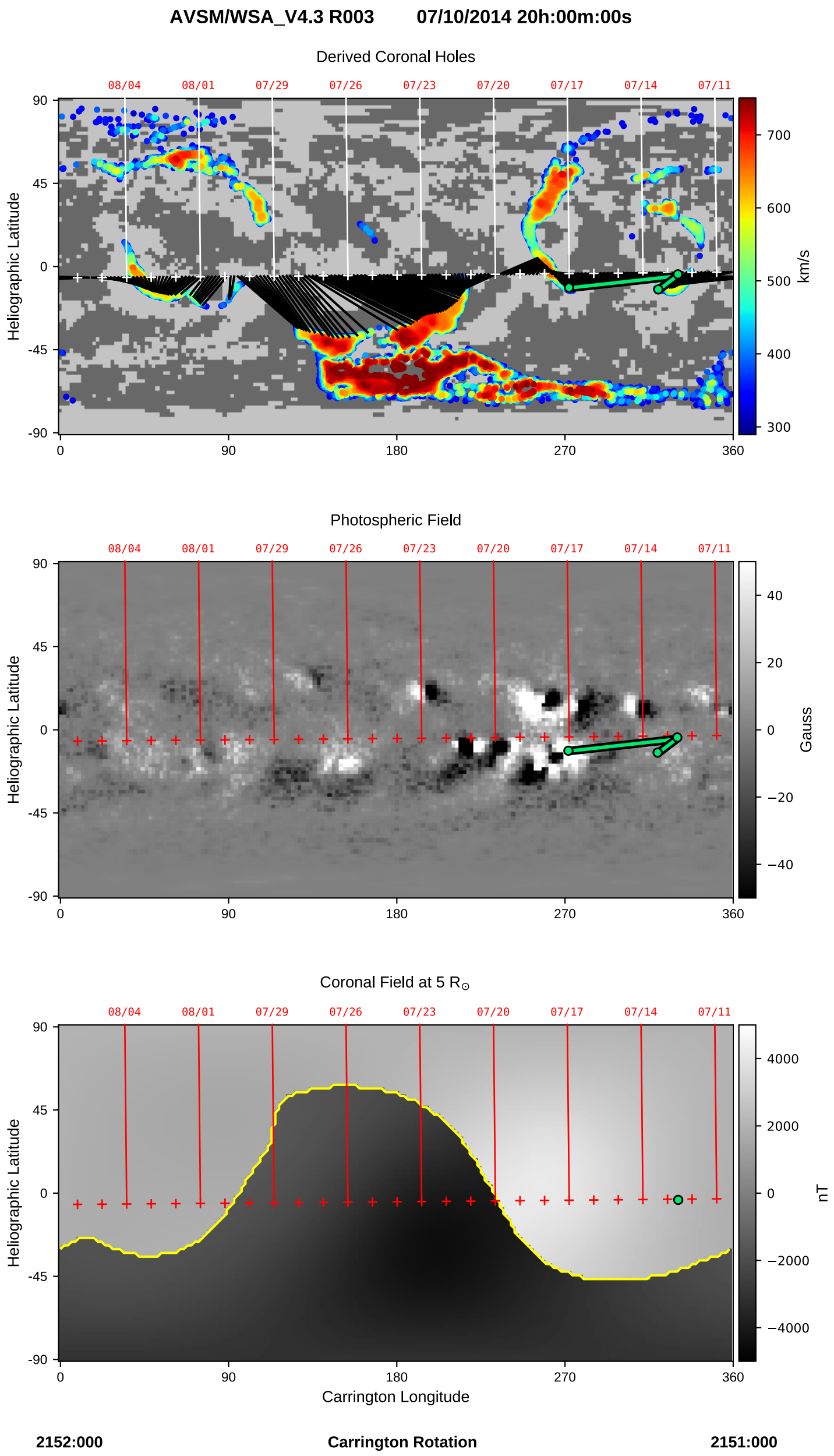

**Figure 1**: ADAPT-WSA model output for CR 2152 (July – August 2014). White (a) or red (b,c) tick-marks label the sub-satellite points, representing the back-projection of STEREO-A's location at 5 $R_\odot$ with dates labeled above in red. **a) (top)** WSA-derived open field at 1 $R_\odot$ with model-derived solar wind speed in colorscale. The field polarity at the photosphere is indicated by the light/dark (positive/negative) gray contours. Black lines show the magnetic connectivity between the projection of STEREO-A's location at 5 $R_\odot$ and solar wind source region at 1 $R_\odot$. Two green lines mark where STEREO-A samples the solar wind that emerged from a pseudostreamer (*i.e.* STEREO-A connectivity changes from one coronal hole boundary to another of the same (outward) polarity). **b) (middle)** Synchronic ADAPT-VSM photospheric field (Gauss) for 07/10/2014 00:00:00 UTC, which reflects the timestamp of the last magnetogram assimilated into this map. See a) for description of two green lines. **c) (bottom)** WSA-derived coronal field at 5 $R_\odot$. Yellow contour marks the model-derived Heliospheric current sheet, where the overall coronal field changes sign.

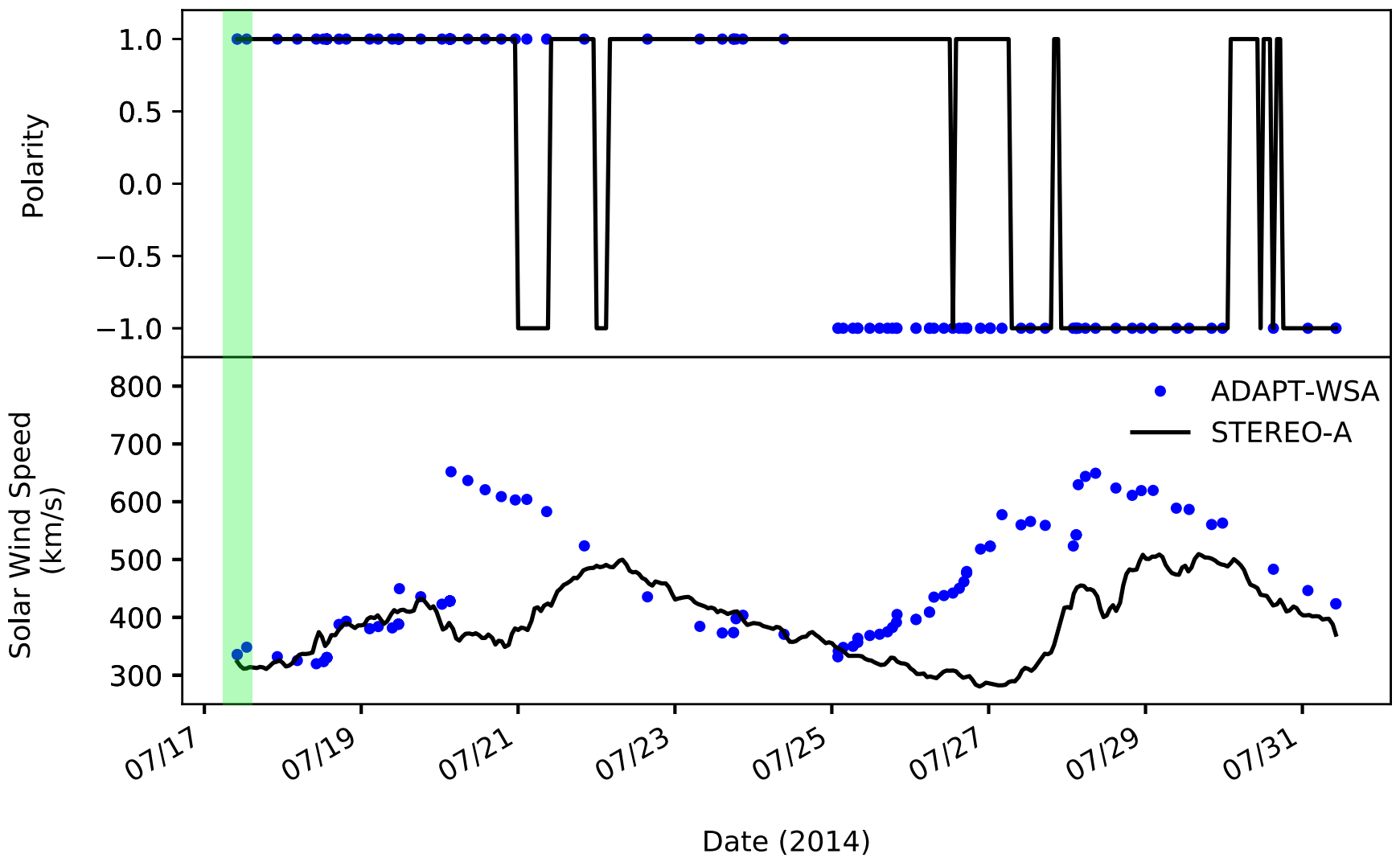

**Figure 2**: ADAPT-WSA model output (blue) vs. STEREO-A observations (black) for approximately two weeks during CR 2152. Each blue dot represents an individual solar wind parcel that connects to STEREO-A, as derived by the model. Highlighted portion in green contains the model output for two solar wind parcels, one that emerged from either last open field line forming the pseudostreamer cusp identified in Figures 1 and 3. **a) (top)** Polarity and **b) (bottom)** Solar wind speed.

We then use a 3D visualization software (GeospaceX 2015) to overplot field line extrapolations of the best ADAPT-WSA realization for each period of interest onto the photosphere (*e.g.* Figure 3). This allows us to identify the specific field lines that the in situ observed pseudostreamer wind emerged from. This software provides the location of each field line footpoint at 1 $R_\odot$ and endpoint at 5 $R_\odot$, allowing us to obtain model-derived parameters specific to each field line such as expansion factor and solar wind speed. Figure 3 shows a subset of open magnetic field lines plotted in triads for CR 2152. The 3D rendering tool allows the user to display this particular subset of field lines, which includes only those magnetically connected to the STEREO-A sub-satellite points at 5 $R_\odot$, and those located a half a grid cell above and below. The middle field lines in each triad correspond to the black lines that map to the white cross hairs (*i.e.* STEREO-A sub-satellite track) in Figure 1a, and are thus the source of each solar wind parcel observed by the spacecraft. Figure 3 shows these field lines as viewed in the ecliptic plane in 3a, and as viewed from the solar North pole in 3b. The two sets of field lines highlighted in green in Figure 3 show where STEREO-A traversed this pseudostreamer, by revealing when STEREO-A changed connectivity from one coronal hole boundary to another of like-polarity. These same field lines correspond to the two green lines in Figure 1a. Once the last open field lines are identified for each pseudostreamer, we obtain the corresponding expansion factors. It is important to note that since we are investigating pseudostreamers observed in situ, observations and field line modeling of each structure only represent a 2D slice of a larger 3D structure.

The model also assigns solar wind speed (Equation 2) to individual solar wind parcels that emerge from each field line along the sub-satellite track (*i.e.* middle field lines in triads Figure 3). These parcels are propagated outward by the model to the observing spacecraft. Their model-determined arrival time is then used to identify the STEREO-A observed solar wind speed. Figure 2 highlights the model-estimated arrival time at STEREO-A of the solar wind that emerged from the pseudostreamer identified in Figures 1 and 3 (all marked in green). Each blue dot in Figure 2 represents the model-derived solar wind speed and IMF for an individual solar wind parcel propagated out from a specific source region (*i.e.* field line of the STEREO-A observed solar wind (Figure 3). The two solar wind parcels that originated from the two last open field lines identified in Figures 1 and 3 (both highlighted in green) were observed at STEREO-A between 10:00:00 - 13:30:00 UTC on July 17th, 2014 as determined by WSA. For this time period, the model-derived and STEREO-observed IMF polarity and solar wind speed are in good agreement. In this study, we did not include periods when WSA produced incorrect polarity or when the model-derived and spacecraft-observed solar wind speeds differ by more than 0.5 days in arrival time. This window is selected because it approximately corresponds to the typical uncertainty of WSA's solar wind speed predictions of ±0.5 days (Owens et al. 2005). We also did not include periods in which coronal mass ejections (CMEs) were identified in situ (Richardson 2014).

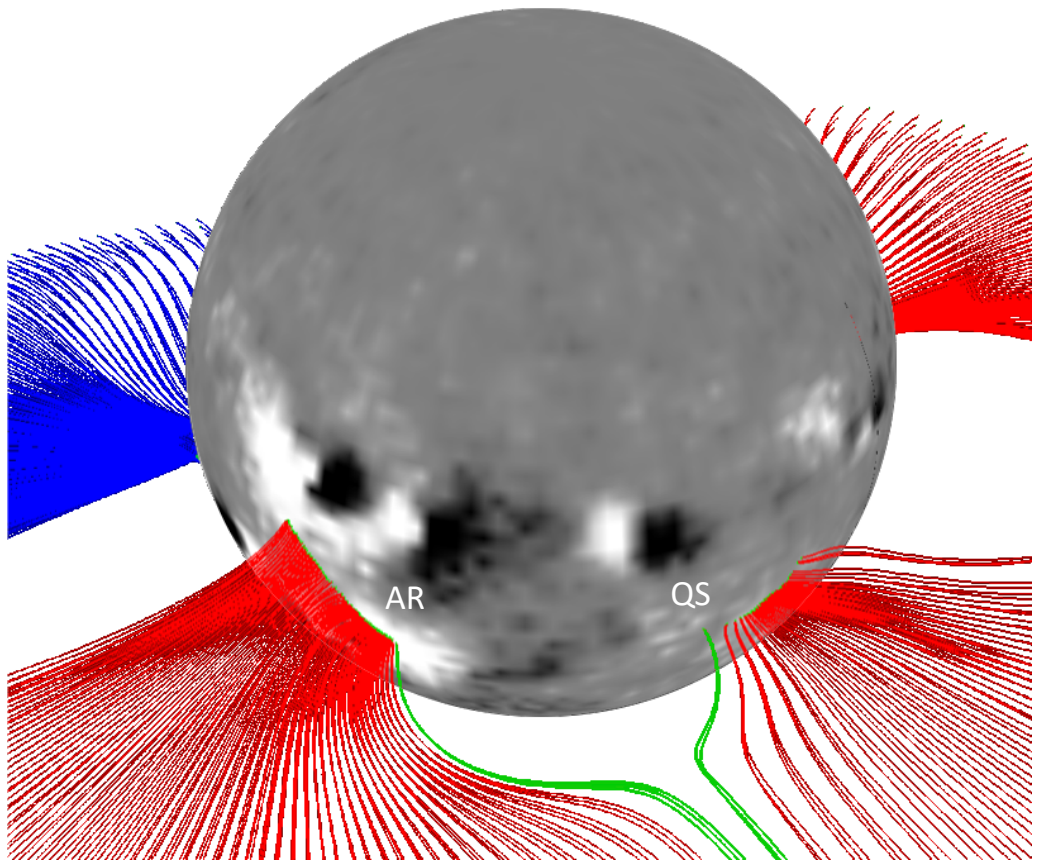

(a)

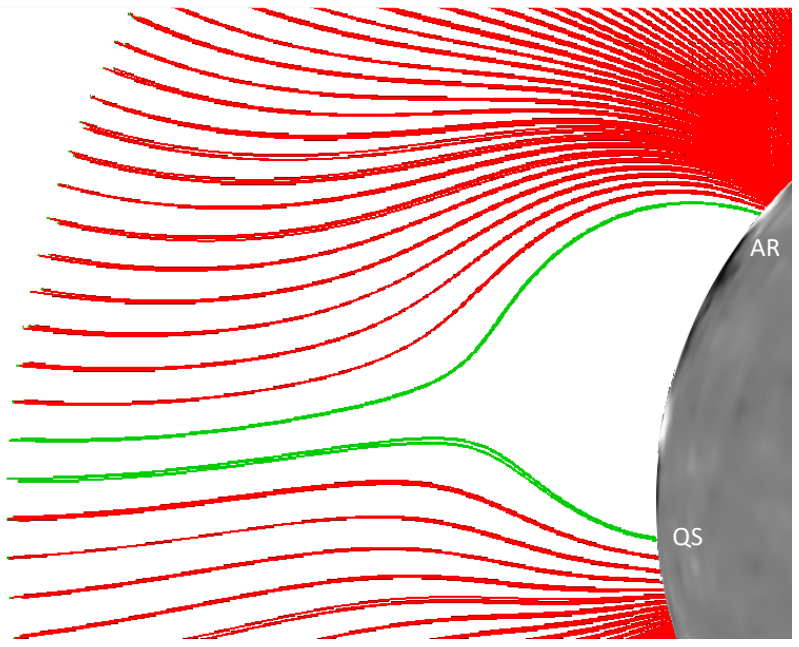

(b)

**Figure 3**: 3D rendering of STEREO-A sub-satellite point magnetic connectivity to 1 $R_{\odot}$ for July 11th - August 6th, 2014. These dates correspond to when the solar wind left the Sun at 5 $R_{\odot}$. Plotted are 180 triads of magnetic field lines at 2° resolution, which include the sources of the observed solar wind at STEREO-A (in the middle of each triad), surrounded by those positioned half a grid cell above and below. Red (blue) lines denote outward (inward) polarity. The type of magnetic field on either side of the pseudostreamer null-point is labeled (*i.e.* "QS" for quiet Sun, "AR" for active region) The last open field lines converging to the pseudostreamer X-point are labeled in green, and viewed from **a) (left)** the ecliptic plane and **b) (right)** the solar North pole.

**Table 1**: Field lines in Figure 3

| | Field line in AR | Field line in QS |
|---|---|---|
| $\lvert B_{ph}\rvert$ (G) | 195.1 | 10.8 |
| $f_s$ | 260.8 | 14.9 |
| $v_{obs}$ (km/s) | 326 | 321 |

Using the above methodology, we investigate over a decade's worth of observations to identify periods when various spacecraft (ACE, STEREO-A/B) observed the solar wind that emerged from pseudsotreamers. We first use diachronic VSM photospheric field maps (*i.e.* one map for each Carrington rotation) as input into WSA to derive the coronal field for CR 2025 − 2185 (Jan. 2005 − Dec. 2016). This provided a quick way to scan through years of model-derived connectivity for the three different spacecraft (*e.g.* Figure 1a), and search for periods when spacecraft could have observed pseudostreamer wind. We then use ADAPT-VSM synchronic maps surrounding each period of interest to drive WSA and produce instantaneous global coronal solutions. In total, we identify 38 unambiguous cases where spacecraft sampled the solar wind that emerged from pseudostreamers.

For each pseudostreamer, we compare the observed solar wind speed and expansion factor for each of the last two open field lines converging at the X-point *separately* (*e.g.* Figure 3, labeled in green). These field lines trace back to two different coronal holes, and their foot points can be grounded in entirely different types of magnetic field back at the Sun – either in quiet Sun (QS) or an active region (AR). An example is shown in Figure 3, where there is asymmetric expansion on either side of the null-point (Figure 3b). Table 1 compares the expansion factor, photospheric field magnitude, and observed solar wind speed for the last open field lines forming this structure (Figure 3, highlighted in green). The order of magnitude difference in magnetic expansion factor is attributed to one field line being rooted in an active region (larger $f_s$), and the other in quiet Sun (smaller $f_s$). By treating each separately, we preserve the expansion history of each field line on either side of the null-point.

## 3. RESULTS

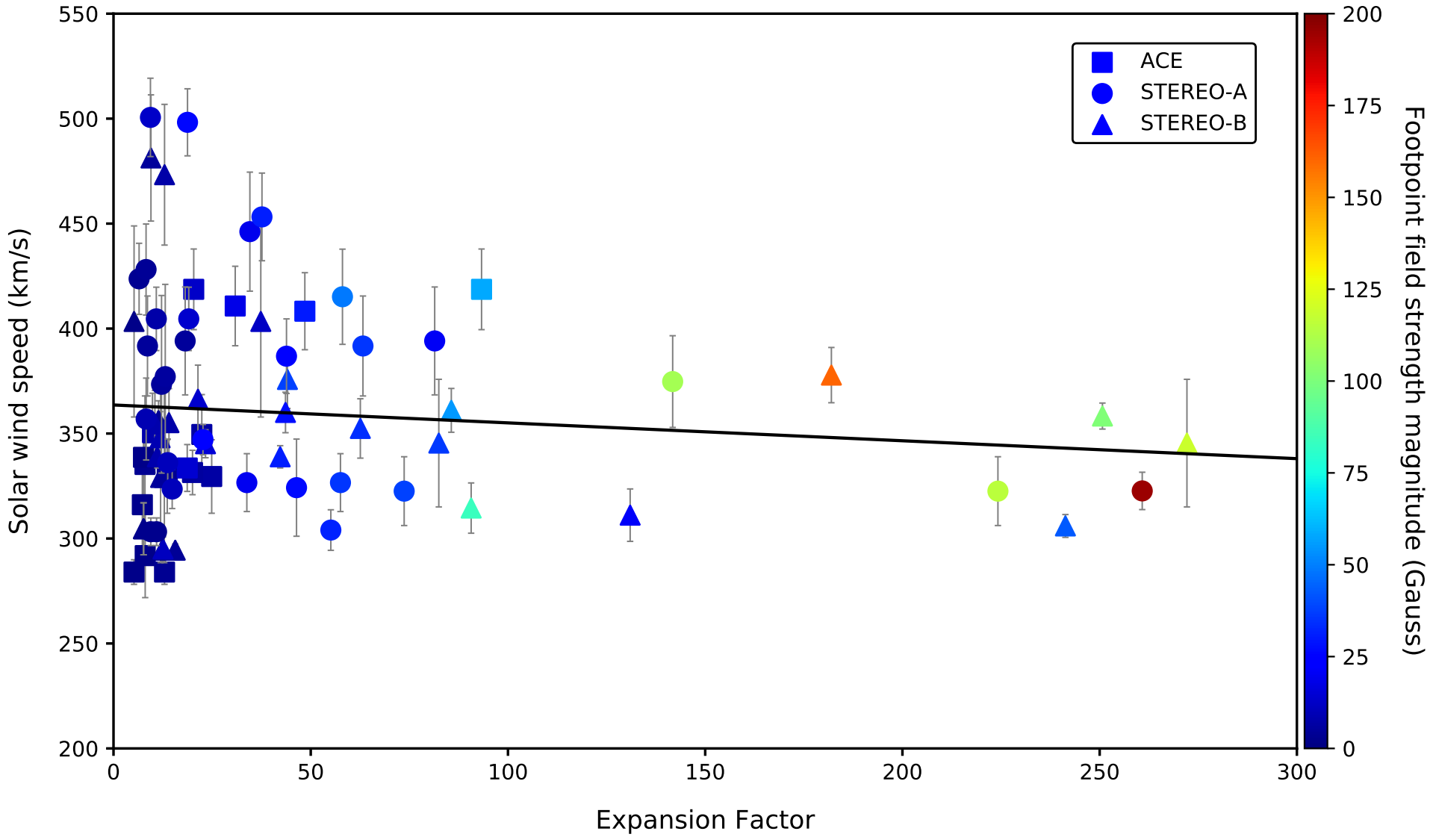

**Figure 4**: Observed solar wind speed vs. expansion factor for all 76 individual last open field lines of the 38 identified pseudostreamers. Each pseudostreamer is represented by two field lines, one from each coronal hole boundary, which together form the 2D slice of a pseudostreamer that was observed by ACE, STEREO-A, or STEREO-B (denoted by different shapes in legend). Observed solar wind speed is averaged over ±0.5 days from the WSA-derived solar wind parcel time-of-arrival at the spacecraft. Error bars in gray represent the standard deviation in observed solar wind speed over a one-day bin, centered on the time (as determined by WSA) that the pseudostreamer-wind was measured at each spacecraft. Black line denotes the calculated linear regression for this dataset.

**Table 2**: Correlation between $f_s$ and $v_{obs}$

| | PCC | $p$-value | No. of field lines |
|---|---|---|---|
| Fig. 4: All Field Lines | -0.1075 | 0.3552 | 76 |
| Fig. 5: QS-QS | 0.2707 | 0.1479 | 30 |
| Fig. 6: AR-AR or AR-QS | -0.3053 | 0.0391 | 46 |

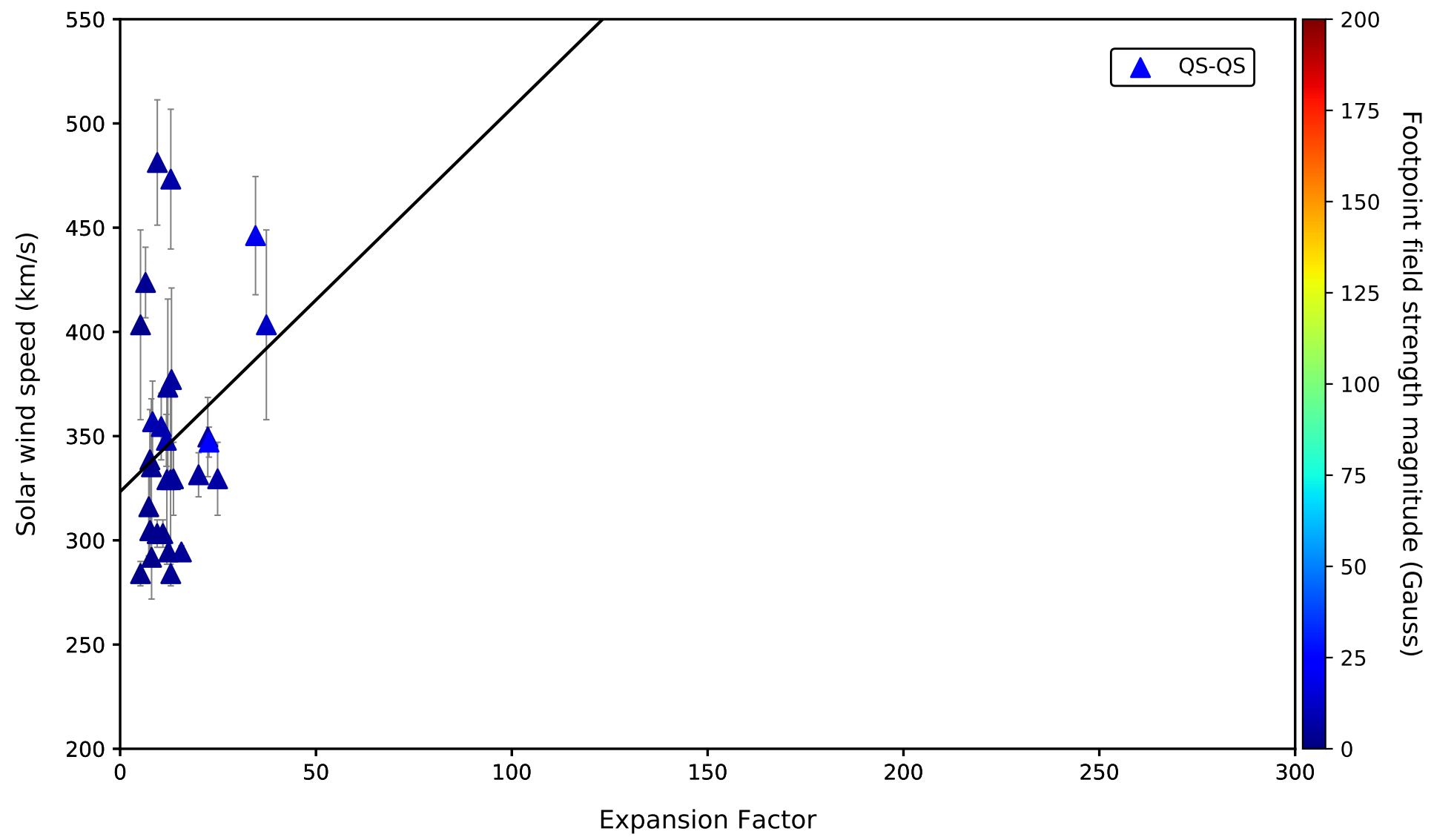

**Figure 5**: Observed solar wind speed vs. expansion factor for 30 individual last open field lines of 15 identified pseudostreamers. Each pseudostreamer is represented by two field lines, one from each coronal hole boundary, and both of which are planted in quiet Sun (QS) photospheric magnetic field, labeled "QS-QS" in the legend. Together, both field lines form the 2D slice of a pseudostreamer that was observed at a spacecraft. Observed solar wind speed is averaged over ±0.5 days from the WSA-derived solar wind parcel time-of-arrival at the spacecraft. Error bars in gray represent the standard deviation in observed solar wind speed over a one-day bin, centered on the time (as determined by WSA) that the pseudostreamer-wind was measured at each spacecraft . Black line denotes the calculated linear regression for this dataset.

Figure 4 compares the observed solar wind speed vs. expansion factor for all individual last open field lines forming the pseudostreamers identified with the methodology outlined in Section 2.2. The observed photospheric field strength at each field line foot point is also shown in colorscale. In Figures 4–7, the observed speed of each solar wind parcel is an average of hourly data over ±0.5 days surrounding when the model-derived parcel arrived at the observing spacecraft. We averaged in this way to account for the ±0.5 day uncertainty window in the model-derived solar wind parcel arrival time, discussed in Section 2.2. Represented in Figure 4 are 38 pseudostreamers and 76 field lines in total (*i.e.* two for each pseudostreamer). This event list spans from the end of solar cycle 23 through most of cycle 24, with pseudostreamers that form at various locations on the disk during both minimum and maximum periods (for a complete list see Appendix). One notable result that is that the solar wind that originates from these field lines is slow ($280 < v_{obs} < 500$ km/s). This is in agreement with prior studies (Wang et al. 2012; Crooker et al. 2012; Riley & Luhmann 2012) which showed that in situ observed pseudostreamer wind is slow for the cases identified in their study. Our results suggest that the observed speed of the solar wind that emerges from these structures is generally slow.

In order to investigate the relationship between expansion factor and speed of the solar wind that emerges from each pseudostreamer field line, we calculated the Pearson correlation coefficient (PCC) for these two quantities and report them in Table 2. Table 2 also lists the $p$-value associated with each correlation coefficient which represents the probability that the correlation occurred at random. The PCC for all field lines in this study (*i.e.* those included in Figure 4) is -0.1075, with approximately a 1 in 3 probability that this result occurred by chance, implying that there is not a statistically significant correlation between $f_s$ and $v_{obs}$ for this dataset. A linear regression was also calculated and included in Figure 4, but it is not likely to have any significance. This result is consistent with prior work (Riley & Luhmann 2012; Riley et al. 2015), which investigated the relationship between $f_s$ and $v_{obs}$ for in situ observed pseudostreamers, but only focused on one period in detail (CR 2060). In these studies, $f_s$-dependent empirical relationships overestimated the observed pseudostreamer solar wind speed. In our study, we use a more robust approach to identify in situ observed pseudostreamer wind over a large, comprehensive sample of periods across the solar cycle.

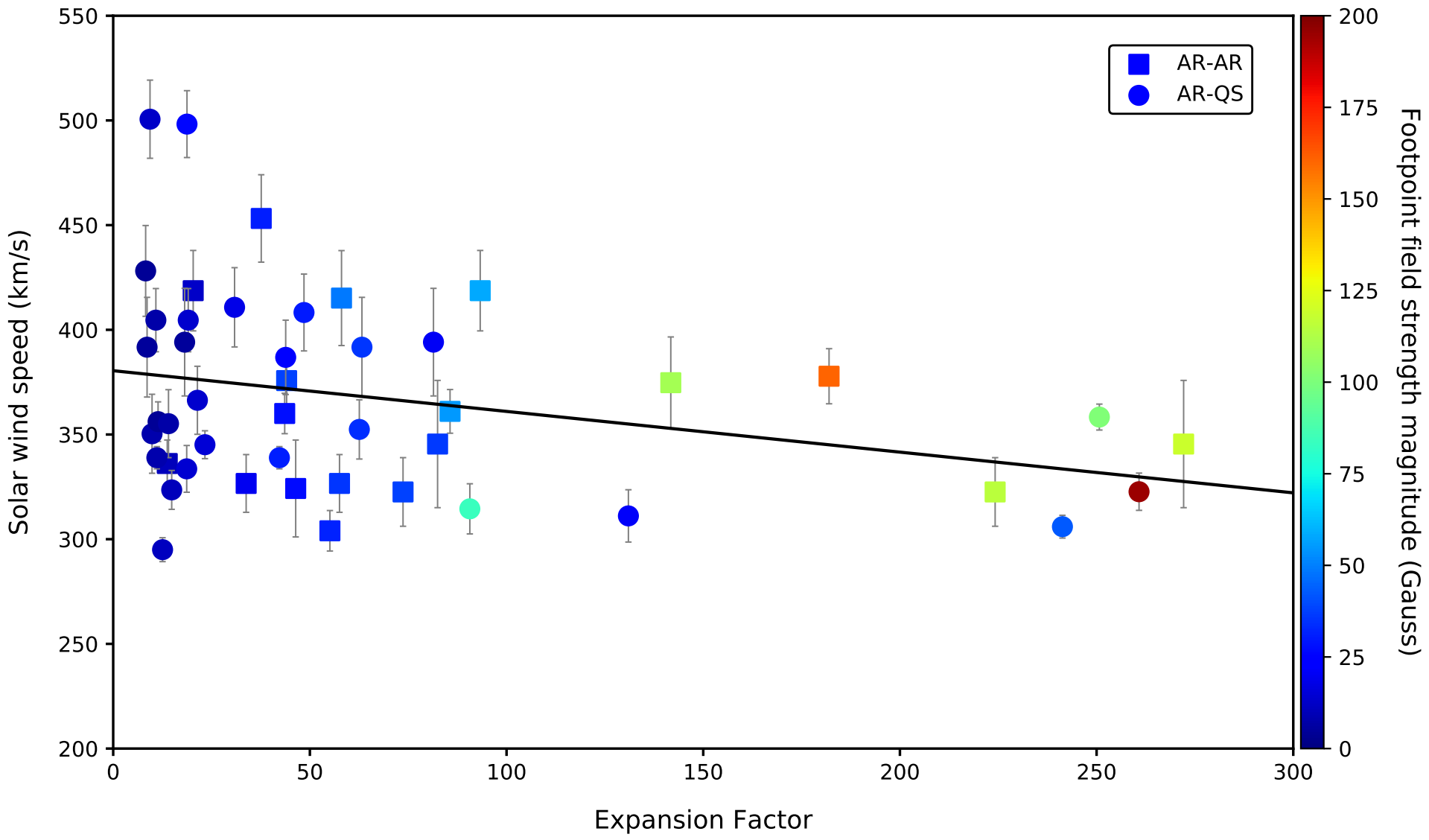

**Figure 6**: Observed solar wind speed vs. expansion factor for 46 individual last open field lines of 23 identified pseudostreamers. Each pseudostreamer is represented by two field lines, one from each coronal hole boundary, where at least one field line is planted in an active region (AR). Field line populations are delineated by shape in the legend, revealing whether an individual field line belongs to a pseudostreamer with one AR and one QS footpoint (AR-QS), or a pseudostreamer with two AR footpoints (AR-AR). Observed solar wind speed is averaged over $\pm 0.5$ days from the WSA-derived solar wind parcel time-of-arrival at the spacecraft. Error bars in gray represent the standard deviation in observed solar wind speed over a one-day bin, centered on the time (as determined by WSA) that the pseudostreamer-wind was measured at each spacecraft . Black line denotes the calculated linear regression for this dataset.

A notable observation from Figure 4 is that field lines rooted in photospheric field of larger magnitude (warmer colors in Figure 4) are always associated with very slow solar wind speeds ($v_{obs} < 400$ km/s). There also appears to be a larger spread in solar wind speed among those field lines rooted in weaker field (*i.e.* $|B_{ph}| < 25$ G). To investigate this, we separated the field lines into two populations based on whether or not at least one field line in each pseudostreamer was planted in an active region. Figures 5 and 6 show the two populations of field lines separated by the pseudostreamer's source region, either entirely quiet Sun (Figure 5) or at least one active region on either side of the cusp (Figure 6). Represented in Figure 5 are 30 of the last open field lines on either side of a pseudostreamer with both footpoints rooted in quiet Sun, labeled "QS-QS" in the legend. For this subset of pseudostreamer field lines, it is apparent that several field lines have similar expansion factors (*i.e.* those with values between $\sim 5-25$), yet the observed speed of the solar wind that emerges from those field lines varies over nearly the entire range of speeds exhibited in this study. The linear regression fit and Pearson correlation coefficient confirm that there is not a statistically significant inverse correlation between observed solar wind speed and expansion factor for these field lines (Table 2).

Figure 6 shows the remaining 46 field lines and is further sub-divided into two populations based on whether each field line is a part of a pseudostreamer with *both* footpoints in an active region (AR-AR), or a pseudostreamer with one footpoint rooted in an active region, and the other footpoint rooted in quiet Sun (AR-QS) These two populations are denoted by different shapes in Figure 6. When considering only those pseudostreamers with at least one field line footpoint planted in an active region, there is now a weak, inverse correlation between $f_s$ and $v_{obs}$ that is approximately at the negligible threshold (*i.e.* $0 \leq |\text{PCC}| < 0.30$ is a negligible correlation). This correlation coefficient in principle is on the borderline of statistical significance because the probability of this result occurring by chance is less than 5% (*i.e.* $p$-value $< 0.05$ is marginally statistically significant, $p$-value $< 0.01$ is statistically significant). However, more would have to be explored to interpret the significance of this result (see Section 4), specifically to understand why there is a weak correlation between $f_s$ and $v_{obs}$ for *only* the pseudostreamers with an active region at at least one of the two coronal hole boundaries.

## 4. DISCUSSION

In this work, we test the original $f_s$–$v_{obs}$ inverse relationship (Wang & Sheeley 1990) by comparing expansion factor and observed speed of the solar wind that emerged from field lines near the open-closed boundary of coronal pseudostreamers. While it is well-established that this relationship can reproduce observed solar wind speed on average over large temporal scales (*i.e.* years, solar cycles), prior work has identified a few examples when in situ observed pseudostreamer wind was much slower than that predicted by $f_s$-dependent empirical relationships. With the methodology outlined in Section 2.2, we identify 38 periods when spacecraft observed the solar wind that emerged from pseudostreamers from 2007 − 2016, and compared the expansion factor of the last open field lines that form these structures with the observed speed of the solar wind.

A significant result of this work is that the observed speed of the solar wind that emerged from all identified pseudostreamers is slow ($v_{obs}$ < 500 km/s). This finding agrees with prior work and strongly suggests that the solar wind that originates from pseudostreamers is slow. Further, in all instances of testing the inverse nature of the original $f_s$–$v_{obs}$ relationship for this dataset, there is not a strong inverse relationship between these two quantities that is statistically significant. When considering only those pseudostreamers with QS field on either side of the cusp (Figure 5), we find that several field lines have the same expansion factor, yet the observed solar wind speed at 1 AU varies over the entire range of speeds exhibited in this study. On the other hand, pseudostreamers that have at least one active region on either side of the cusp exhibit a weak, inverse relationship between $f_s$ and $v_{obs}$ that is marginally statistically significant (*i.e.* $p$-value is less than 5%). However, the distribution of speed versus expansion factor seen in Figures 4 − 6 exhibits an envelope such that when $f_s$ is small, a large range of speeds are observed, whereas when $f_s$ is large, only slow wind speeds are observed. This may indicate that expansion factor plays a role in setting an upper threshold on the observed solar wind speed.

An interesting finding of this study is the asymmetric global magnetic topology of a pseudostreamer that results from differences in the local magnetic topology at the field line footpoints of the two converging coronal hole boundaries. An example is shown in Figure 3 and Table 1, where there is an active region at one coronal hole boundary, and quiet Sun at the other. We also found that field lines grounded in the largest magnitude of photospheric field are also associated with slower solar wind on average ($v_{obs}$ < 400 km/s). One possibility is that the in situ observed properties of the solar wind are more dependent on the local magnetic field at the field line footpoint (*e.g.* active region vs. quiet Sun) than the global magnetic structure (*e.g.* pseudostreamer vs. helmet streamer). This hypothesis will be tested in future work and could have implications for solar wind formation theories.

While these results rigorously substantiate prior pseudostreamer case studies that have small sample sizes, it is important to note that the conclusions of this work are only applicable to pseudostreamers. Thus, we do not conclude that flux-tube expansion plays no physical role in solar wind acceleration. This follows for several reasons. First, expansion factor is calculated in this study as the rate of flux tube expansion from 1 to 2.5 $R_\odot$, as originally defined by Wang & Sheeley (1990). However, Panasenco & Velli (2013) argue that expansion factor as originally defined is not appropriate for pseudostreamers because field line expansion does not increase monotonically with distance from the Sun as in helmet streamers. They propose the use of a 3D calculation of expansion factor to predict solar wind speed for pseudostreamers, arguing that this quantity better captures the entire magnetic field configuration (*e.g.* height of X-point, separation between corona holes). Second, recent work by Wang & Panasenco (2019) employ the use of the maximum value of expansion factor ($f_{max}$) along a field line, as opposed to quantifying field line expansion from 1 − 2.5 $R_\odot$ (Equation 1), in an empirical relationship to determine $v_{obs}$ for ten different pseudostreamers observed at L1. Their results suggest that using $f_{max}$ in lieu of $f_s$ recovers the inverse relationship between speed and field line expansion for these pseudostreamers, though they argue that a single 2D parameter cannot fully describe the non-monotonic expansion along the last open field lines forming pseudostreamers.

It is possible that using the traditional definition of expansion factor ($f_s$) vs. the maximum value ($f_{max}$) in our study could explain the weak, inverse correlation in Figure 6, and the lack of a correlation in Figure 5. Field lines with stronger photospheric fields at their base undergo more expansion from 1 to 2.5 $R_\odot$ which is captured by $f_s$. However, pseudostreamer field lines rooted in quiet Sun converge much lower down, and thus their overall expansion is not well-captured by the traditional definition of expansion factor (Equation 1). A means of testing this theory would be to reproduce this study (*i.e.* Figures 4–6) using $f_{max}$, to see if a stronger, inverse relationship exists between $f_{max}$ and $v_{obs}$ that is statistically significant. However, it remains to be shown if replacing $f_s$ with $f_{max}$ in either the original WS prescription or the updated WSA relationship (Equation 2) would reproduce the observed speed of the solar wind originating from 1) other magnetic sources (*e.g.* helmet streamers, coronal holes), and 2) on large temporal scales as originally demonstrated (Wang & Sheeley 1990, Figure 3).

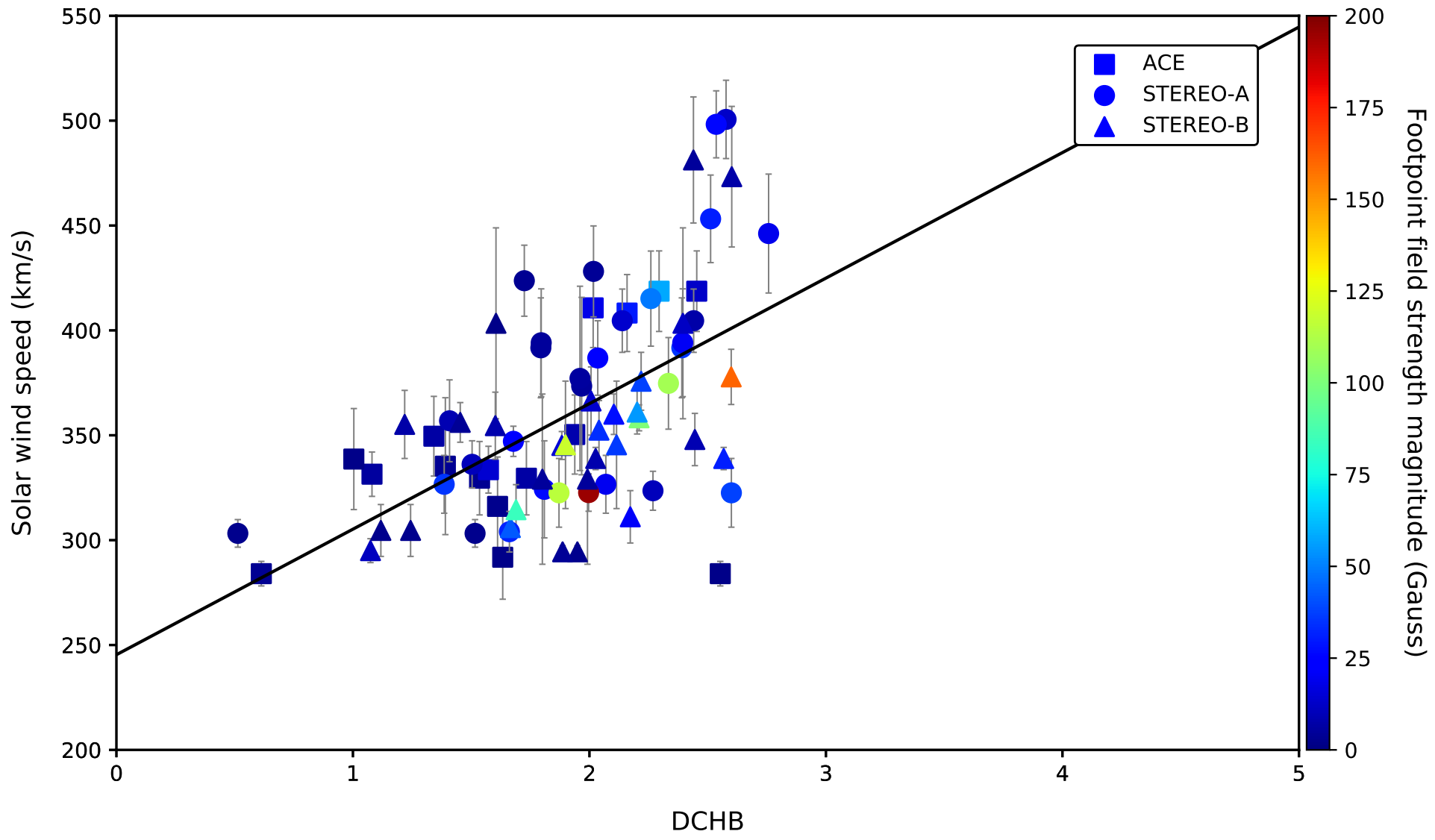

**Figure 7**: As in Figure 4, but comparing observed solar wind speed with the model-derived DCHB.

Third, we are comparing the observed speed of the solar wind that emerged from the last open model-derived field lines converging at a null point within the pseudostreamer. These field lines form the open-closed boundary of their respective coronal holes – an ideal environment for interchange reconnection. If interchange reconnection is an ubiquitous, time-dependent effect at the open-closed boundary, then we would expect any signature of an inverse $f_s$–$v_{obs}$ relationship to be undetectable as flux is constantly opening up and closing down. Future work will investigate this by testing if the inverse $f_s$–$v_{obs}$ correlation is recovered for solar wind that 1) *emerged* from pseudostreamer field lines farther away from open-closed boundary (*e.g.* the $2-3$ red field lines surrounding those in green in Figure 3), or 2) the last open field lines forming helmet streamers.

Figure 7 probes the role possibly played by interchange reconnection for this set of field lines by comparing $\theta_b$ of each pseudostreamer field line and $v_{obs}$ of each corresponding solar wind parcel observed in situ. Although our capacity to test this with high fidelity is limited because the grid resolution of the model is coarser than the computed resolution of $\theta_b$, Figure 7 suggests there is a possible relationship between these two variables. First, the PCC of this dataset is 0.5670, with a $p$-value of $10^{-7}$, meaning that there is a moderate positive correlation with high statistical significance ($p$-value $< 0.01$), with an extremely low probability of occurring by chance. This plot includes all of the pseudostreamer field lines in this study, whereas when the same comparison was made between $f_s$ and $v_{obs}$ (*e.g.* Figure 4), there was no correlation whatsoever. Second, although there are essentially two bins in this plot, $0° \leq \theta_b < 2°$ and $2° \leq \theta_b < 4°$, within those bins we see that there is a smaller spread of observed speeds and lower average $v_{obs}$ from $0° \leq \theta_b < 2°$. This result suggests that only the slowest wind is observed at the open-closed boundary ($\theta_b \sim 0-2°$), and solar wind that emerged farther away from this boundary (as $\theta_b$ increases) exhibits a wider range of speed. It's possible that magnetic field near the open-closed boundary but not quite deep inside a coronal hole ($\theta_b \sim 2-3°$) is a mixture of continuously open field and flux tubes that are intermittently open due to interchange reconnection, resulting in a wider speed range of solar wind that emerges from this region. These preliminary conclusions will be tested in future work with 1° model resolution runs.

## 5. SUMMARY

In this work, we test the original $f_s$–$v_{obs}$ inverse relationship (Wang & Sheeley 1990) by performing a statistical analysis comparing expansion factor and observed speed of the solar wind that emerged from field lines near the open-closed boundary of coronal pseudostreamers. We exploit new advances in the ADAPT-WSA model to develop a methodology to more rigorously determine the precise source region of the in situ observed solar wind. This methodology has already been used to interpret the first observations from Parker Solar Probe (Szabo et al. 2020; Korreck et al. 2020; Nieves-Chinchilla et al. 2020; Hill et al. 2020), and is extremely useful for coordinated multi-messenger science between remote coronal and in situ solar wind observatories.

Using ADAPT-WSA, we identify 38 periods where either ACE, STEREO-A, or STEREO-B sample the solar wind that emerged from pseudostreamers. This study is the first to identify a large sample of in situ observed pseudostreamers with multiple spacecraft, to more thoroughly investigate whether $f_s$-dependent empirical relationships perform poorly when pseudostreamer wind is observed in situ as suggested by prior case studies (Riley & Luhmann 2012; Riley et al. 2015). For the 38 pseudostreamers we identified, the observed solar wind speed ranges from $\sim 280 - 500$ km/s suggesting that pseudostreamer wind is slow on average. We also find that there is not a statistically significant correlation between $f_s$ and $v_{obs}$ for solar wind that emerged near the open-closed boundary of pseudostreamers. This result is somewhat expected, considering we are investigating field lines near the magnetic open-closed boundary where it is likely that flux tubes are intermittently open due to interchange reconnection. Since this work does not address the vast majority of solar wind outflow along continuously open field lines, it's possible that flux tube expansion, regardless of how it is quantified (*e.g.* $f_s$, $f_{max}$, 3D expansion factor), could still play an important role in modulating solar wind speed for those field lines that are continuously open (*i.e.* deeper inside the coronal hole). If this were the case, it could explain the wider speed range exhibited as $\theta_b$ increases in Figure 7. This hypothesis will be tested in future work.

Acknowledgments: This work was supported by NASA (Grant No. 80NSSC17K0606 P00003) and the Air Force Research Lab (Grant No. FA9453-15-1-0333). NMV is supported by the Heliophysics Internal Scientist Funding Model. This work utilizes ADAPT maps produced collaboratively between AFRL and NSO/NISP. NSO/Kitt Peak data used here are produced cooperatively by NSF/NSO, NASA/GSFC, and NOAA/SEL. SOLIS data for this work are obtained and managed by NSO/NISP, operated by AURA, Inc. under a cooperative agreement with NSF. In situ measurements were obtained from NASA/GSFCs Coordinated Data Analysis Web (CDAweb - https://cdaweb.gsfc.nasa.gov/index.html/).

## APPENDIX

## A. PSEUDOSTREAMER TABLES

Below are tables listing all ADAPT-WSA derived pseudostreamer field lines used in this work. Each pseudostreamer is represented by two field lines, one from each coronal hole boundary that converges to form a pseudostreamer X-point. These two field lines mark when the spacecraft connectivity changes from one side of the pseudostreamer cusp to the other (*i.e.* from one coronal hole boundary to another of like-polarity). In total there are 38 pseudostreamers observed, and 76 individual field lines that are the sources of solar wind observed at either ACE, STEREO-A, or STEREO-B. The following tables also list the solar wind parcel observation times at each spacecraft that correspond to a particular field line.

Pseudostreamers are assigned an alphanumeric label in the first column of each table. Since there are two field lines associated with each pseudostreamer, the first two rows of each table and every proceeding pair of rows share the same alphanumeric label. Each unique pseudostreamer is first assigned a number. If this same structure has either been observed at a different spacecraft, or resampled by the same spacecraft in another rotation, it is then marked with a letter proceeding the reference number. In either scenario, the pseudostreamer undergoes evolution and the spacecraft sample a different 2D slice of the 3D structure. For example, a label of 1a and 1b would be two different samples in time and space of the "same" pseudostreamer.

**Table 3**: Pseudostreamers observed at ACE

| Label[a] | CR | Footpoint coords. (Lat., Carr. Long.) | Field at 1 $R_\odot$ (AR or QS) | Date SW[b] observed (yyyy-mm-dd) | s/c time of arrival (hh:mm:ss) | Polarity |
|---|---|---|---|---|---|---|
| *ACE* | | | | | | |
| 1a | 2060 | (-0.09, 345.92) | QS | 2007-08-21 | 13:19:29 | inward |
| 1a | 2060 | (70.73, 340.86) | QS | 2007-08-21 | 13:19:38 | inward |
| 2a | 2060 | (26.46, 267.77) | QS | 2007-08-25 | 05:16:31 | inward |
| 2a | 2060 | (6.22, 257.84) | QS | 2007-08-25 | 11:18:49 | inward |
| 1b | 2061 | (-2.31, 354.73) | QS | 2007-09-16 | 22:37:21 | inward |
| 1b | 2061 | (65.77, 356.81) | QS | 2007-09-17 | 12:25:47 | inward |
| 1c | 2062 | (0.55, 6.57) | QS | 2007-11-08 | 11:08:27 | inward |
| 1c | 2062 | (0.25, 5.67) | QS | 2007-11-09 | 06:14:07 | inward |
| 3a | 2075 | (61.61, 147.01) | QS | 2008-10-18 | 04:18:03 | inward |
| 3a | 2075 | (33.86, 132.12) | QS | 2008-10-18 | 04:18:12 | inward |
| 4a | 2109 | (-14.07, 282.85) | QS | 2011-04-22 | 20:02:33 | outward |
| 4a | 2109 | (11.67, 261.09) | AR | 2011-04-22 | 21:46:22 | outward |
| 5a | 2109 | (14.65, 258.64) | AR | 2011-04-24 | 06:34:42 | outward |
| 5a | 2109 | (-18.30, 220.52) | AR | 2011-04-24 | 06:34:51 | outward |
| 6 | 2164 | (10.74, 77.86) | AR | 2015-06-19 | 15:48:49 | outward |
| 6 | 2164 | (11.05, 55.40) | QS | 2015-06-20 | 01:19:03 | outward |

[a] Each unique pseudostreamer is assigned a number in this table. If a pseudostreamer was observed at more than one spacecraft, or observed in another rotation (even if several rotations ahead), a letter is assigned. In some cases the same pseudostreamer is observed several rotations later after significant evolution has occurred.

[b] Solar wind.

**Table 4**: Pseudostreamers observed at STEREO-A

| Label[a] | CR | Footpoint coords. (Lat./Carr. Long.) | Field at 1 $R_\odot$ (AR or QS) | Date SW[b] observed (yyyy-mm-dd) | s/c time of arrival (hh:mm:ss) | Polarity |
|---|---|---|---|---|---|---|
| *STEREO-A* | | | | | | |
| 2b | 2060 | (50.97, 250.96) | QS | 2007-08-26 | 12:11:23 | inward |
| 2b | 2060 | (-2.89, 241.52) | QS | 2007-08-26 | 13:05:14 | inward |
| 2c | 2062 | (61.11, 261.96) | QS | 2007-10-18 | 07:46:51 | inward |
| 2c | 2062 | (18.44, 251.51) | QS | 2007-10-18 | 07:47:00 | inward |
| 7a | 2100 | (33.93, 154.84) | QS | 2010-09-05 | 09:55:09 | inward |
| 7a | 2100 | (25.66, 108.46) | AR | 2010-09-05 | 09:55:26 | inward |
| 8 | 2101 | (-13.65, 193.77) | QS | 2010-09-29 | 11:51:30 | inward |
| 8 | 2101 | (14.61, 180.37) | QS | 2010-09-30 | 13:22:48 | inward |
| 9a | 2109 | (13.29, 315.38) | AR | 2011-04-29 | 01:39:39 | outward |
| 9a | 2109 | (7.51, 273.23) | AR | 2011-04-29 | 20:09:01 | outward |
| 5b | 2109 | (14.53, 258.35) | AR | 2011-05-03 | 20:38:15 | outward |
| 5b | 2109 | (-18.14, 219.19) | AR | 2011-05-03 | 20:38:24 | outward |
| 9b | 2110 | (18.69, 312.88) | AR | 2011-05-27 | 04:54:20 | outward |
| 9b | 2110 | (15.51, 260.24) | AR | 2011-05-27 | 14:15:30 | outward |
| 10 | 2110 | (12.88, 127.30) | AR | 2011-06-10 | 03:31:58 | inward |
| 10 | 2110 | (37.59, 80.02) | QS | 2011-06-10 | 03:32:07 | inward |
| 11a | 2112 | (0.12, 168.82) | QS | 2011-08-01 | 00:55:44 | inward |
| 11a | 2112 | (-20.44, 116.91) | AR | 2011-08-01 | 14:12:55 | inward |
| 11b | 2113 | (3.36, 168.76) | QS | 2011-08-28 | 15:45:56 | inward |
| 11b | 2113 | (-22.99, 113.56) | AR | 2011-08-28 | 15:46:05 | inward |
| 12a | 2113 | (17.10, 64.95) | AR | 2011-09-04 | 10:53:28 | inward |
| 12a | 2113 | (19.18, 28.26) | AR | 2011-09-04 | 10:53:37 | inward |
| 13 | 2116 | (-14.65, 95.57) | AR | 2011-11-21 | 16:01:12 | inward |
| 13 | 2116 | (8.99, 65.91) | AR | 2011-11-22 | 05:23:43 | inward |
| 14 | 2136 | (29.35, 244.20) | QS | 2013-05-11 | 15:03:27 | inward |
| 14 | 2136 | (20.47, 212.81) | AR | 2013-05-11 | 19:44:41 | inward |
| 15a | 2136 | (28.64, 123.39) | QS | 2013-05-23 | 10:22:39 | outward |
| 15a | 2136 | (30.18, 102.12) | QS | 2013-05-23 | 19:24:14 | outward |
| 16 | 2152 | (-9.12, 320.18) | QS | 2014-07-17 | 10:03:04 | outward |
| 16 | 2152 | (-10.99, 273.15) | AR | 2014-07-17 | 12:56:44 | outward |

[a] Each unique pseudostreamer is assigned a number in this table. If a pseudostreamer was observed at more than one spacecraft, or observed in another rotation (even if several rotations ahead), a letter is assigned. In some cases the same pseudostreamer is observed several rotations later after significant evolution has occurred.

[b] Solar wind.

**Table 5**: Pseudostreamers observed at STEREO-B

| Label[a] | CR | Footpoint coords. (Lat./Carr. Long.) | Field at 1 $R_\odot$ (AR or QS) | Date SW[b] observed (yyyy-mm-dd) | s/c Time of Arrival (hh:mm:ss) | Polarity |
|---|---|---|---|---|---|---|
| *STEREO-B* | | | | | | |
| 2d | 2060 | (50.79, 250.65) | QS | 2007-08-24 | 12:52:25 | inward |
| 2d | 2060 | (-2.89, 243.63) | QS | 2007-08-24 | 12:52:34 | inward |
| 2e | 2064 | (57.53, 290.82) | QS | 2007-12-08 | 22:36:03 | inward |
| 2e | 2064 | (14.02, 260.21) | QS | 2007-12-08 | 22:36:12 | inward |
| 2f | 2065 | (60.95, 322.81) | QS | 2008-01-02 | 20:41:51 | inward |
| 2f | 2065 | (13.70, 325.59) | QS | 2008-01-02 | 20:42:00 | inward |
| 3b | 2076 | (32.77, 119.62) | AR | 2008-11-15 | 09:21:19 | inward |
| 3b | 2076 | (56.09, 112.70) | QS | 2008-11-15 | 20:50:30 | inward |
| 7b | 2100 | (31.92, 157.94) | QS | 2010-08-25 | 21:49:32 | inward |
| 7b | 2100 | (24.76, 108.77) | AR | 2010-08-25 | 23:13:47 | inward |
| 7c | 2101 | (29.67, 145.69) | QS | 2010-09-20 | 00:46:39 | inward |
| 7c | 2101 | (20.42, 114.29) | QS | 2010-09-20 | 02:24:00 | inward |
| 17 | 2107 | (26.27, 167.79) | AR | 2011-03-01 | 21:24:55 | inward |
| 17 | 2107 | (37.54, 150.30) | AR | 2011-03-02 | 01:15:53 | inward |
| 18 | 2107 | (20.93, 135.75) | AR | 2011-03-04 | 07:56:04 | inward |
| 18 | 2107 | (16.66, 111.79) | QS | 2011-03-04 | 07:56:12 | inward |
| 4b | 2109 | (-12.97, 283.38) | QS | 2011-04-15 | 23:47:02 | outward |
| 4b | 2109 | (11.17, 262.91) | AR | 2011-04-16 | 19:48:35 | outward |
| 11c | 2113 | (-0.48, 169.62) | QS | 2011-08-13 | 23:44:10 | inward |
| 11c | 2113 | (-24.79, 115.26) | QS | 2011-08-13 | 23:44:18 | inward |
| 12b | 2113 | (16.55, 64.93) | AR | 2011-08-20 | 15:54:52 | inward |
| 12b | 2113 | (19.01, 28.24) | AR | 2011-08-20 | 15:55:00 | inward |
| 19a | 2120 | (35.03, 289.95) | QS | 2012-02-09 | 08:14:47 | inward |
| 19a | 2120 | (18.46, 238.15) | AR | 2012-02-09 | 12:31:58 | inward |
| 19b | 2121 | (21.64, 289.26) | AR | 2012-03-07 | 03:13:24 | inward |
| 19b | 2121 | (19.15, 242.78) | AR | 2012-03-07 | 05:41:25 | inward |
| 15b | 2138 | (-13.51, 132.65) | AR | 2013-06-25 | 09:23:20 | outward |
| 15b | 2138 | (30.07, 101.06) | QS | 2013-06-26 | 04:03:13 | outward |
| 15c | 2139 | (16.99, 54.44) | QS | 2013-07-23 | 22:07:32 | outward |
| 15c | 2139 | (11.32, 53.40) | QS | 2013-07-24 | 09:48:40 | outward |

[a] Each unique pseudostreamer is assigned a number in this table. If a pseudostreamer was observed at more than one spacecraft, or observed in another rotation (even if several rotations ahead), a letter is assigned. In some cases the same pseudostreamer is observed several rotations later after significant evolution has occurred.

[b] Solar wind.